\documentclass[a4paper]{jpconf}
\usepackage{graphicx}
\usepackage{epsf}
\usepackage[]{latexsym}
\usepackage{bm}
\newcommand{\be}{\begin{equation}}\newcommand{\ee}{\end{equation}}
\newcommand{\bea}{\begin{eqnarray}}\newcommand{\eea}{\end{eqnarray}}
\newcommand{\brr}{\begin{array}}\newcommand{\err}{\end{array}}
\newcommand{\bit}{\begin{itemize}}\newcommand{\eit}{\end{itemize}}
\newcommand{\ben}{\begin{enumerate}}\newcommand{\een}{\end{enumerate}}

\newcommand{\ba}{\begin{array}}
\newcommand{\ea}{\end{array}}

\def\noi{\noindent}

\def\De{\Delta}

\def\1{{_{1}}}\def\2{{_{2}}}

\def\noHe0{:\;\!\!\;\!\!:H_e(0):\;\!\!\;\!\!:}
\def\noHm0{:\;\!\!\;\!\!:H_\mu(0):\;\!\!\;\!\!:}

\def\noi{\noindent}

\def\De{\Delta}

\def\1{{_{1}}}\def\2{{_{2}}}

\begin{document}
\title{The deformation parameter of the generalized uncertainty principle}

\author{Fabio Scardigli$^{\dagger *}$}

\address{
$^\dagger$ Dipartimento di Matematica, Politecnico di Milano, Piazza Leonardo da Vinci 32, 20133 
\\$\ $ Milano, Italy\\
$^{*}$ Institute-Lorentz for Theoretical Physics, Leiden University, P.O.~Box 9506, Leiden, The 
\\ $\ $ Netherlands}

\ead{fabio@phys.ntu.edu.tw}

\begin{abstract}
After a short introduction to the generalized uncertainty principle (GUP), we review some of the physical predictions of the GUP, and we focus in particular on the bounds that present experimental tests can put on the value of the deformation parameter $\beta$. We also describe a theoretical value  computed for $\beta$, and comment on the vast parameter region still unexplored, and to be probed by future experiments.    
\end{abstract}


\section{Introduction}
\vspace{2mm}

Uncertainty Principle (HUP) emerged when Heisenberg~\cite{Heisenberg1}, in a seminal paper of 1927, discussed measurement processes in Quantum Theory. The argument (for example, find the position of an electron by means of photons) is well known under the name of \textit{Heisenberg microscope argument}~\cite{Heisenberg2}. In those early approaches the gravitational interaction between particles was completely neglected, although this was somehow justified by the huge weakness of gravity, when compared with other fundamental interactions. Then, in a few years, HUP became a theorem at the core of the formalism of a fully developed Quantum Mechanics~\cite{Robertson}.     

However, when basic principles and elementary measurement processes are discussed in order to address fundamental questions in Nature, it seems clear that also gravity should be considered. 
This is what actually happened in the subsequent decades, starting from the very early attempts in generalizing HUP~\cite{GUPearly}, to the more recent proposals like those of string theory, deformed special relativity, and studies of black hole 
physics~\cite{VenezGrossMende,MM,KMM,FS,Adler2,SC-CQG,CGS,SC2013,ADV}.

Various revised versions of the classical Heisenberg argument have been presented, and, for example in Ref.~\cite{FS}, one of them is described as follows. A beam of photons of energy $E$ can theoretically detect an object of size $\delta x$ roughly given by (if we assume
the dispersion relation $E=pc$)
\be
\delta x
\simeq
\frac{\hbar c}{2\, E}
\ ,
\label{HS}
\ee
so that increasingly large energies allow to explore decreasingly small details.
As remarked above, in its original formulation, Heisenberg's gedanken experiment ignores gravity.
Now, if on the contrary in this gedanken measurement process we take into account \textit{also} the possible formation, in high energy scatterings, of micro black holes with a
gravitational radius $R_S=R_S(E)$ roughly proportional to the (centre-of-mass) scattering energy $E$ 
(see Ref.~\cite{FS}), then it turns out that the usual uncertainty relation should be modified as
\be
\delta x
\ \simeq \
\frac{\hbar c}{2\, E}
\ + \
\beta\, R_S(E)
\ ,
\ee
where $\beta$ is a dimensionless parameter.
Recalling that $R_S\simeq 2\,G_N\,E/c^4 = 2\, \ell_p^2\, E/\hbar c$, we can write
\be
\delta x
\ \simeq \
\frac{\hbar c}{2\, E}
\ + \
\beta\,\ell_p^2\,\,\frac{2\, E}{\hbar c} \ ,
\label{He}
\ee
where the Planck length is defined as $\ell_p^2=G\hbar/c^3$, the Planck energy as 
$E_p\ell_p=\hbar c/2$, the Planck mass $m_p=E_p/c^2$.
This kind of modification of the uncertainty principle was also proposed in Ref.~\cite{Adler2}.
\par
Although the deforming parameter $\beta$ is not in principle fixed by the theory, it is generally assumed to be of the order of unity. This happens, in particular, in some models of string theory (see again for instance Ref.~\cite{VenezGrossMende}). An explicit analytic calculation of $\beta$ in Ref.~\cite{SLV} has confirmed the circumstance. 
However, many studies have appeared in literature, with the aim to set experimental bounds on $\beta$
(see, for instance, Refs.~\cite{brau}).
\par
The relation~(\ref{He}) can be recast in the form of an uncertainty relation, namely a deformation
of the standard HUP, usually referred to as Generalized Uncertainty Principle (GUP),
\be
\Delta x\, \Delta p
\geq
\frac{\hbar}{2}
\left[1
+\beta
\left(\frac{\Delta p}{m_p c}\right)^2
\right]
\ .
\label{gup}
\ee
For mirror-symmetric states (with $\langle \hat{p} \rangle = 0$), the inequality~(\ref{gup}) is equivalent to the commutator
\be
\left[\hat{x},\hat{p}\right]
=
i \hbar \left[1 + \beta \left(\frac{\hat{p}}{m_p c} \right)^2 \right]\ ,
\label{gupcomm}
\ee
since $\Delta x\, \Delta p \geq (1/2)\left|\langle [\hat{x},\hat{p}] \rangle\right|$.
Vice-versa, the commutator~(\ref{gupcomm}) implies the inequality~(\ref{gup}) for any state.
The GUP is widely studied in the context of quantum mechanics~\cite{Pedram},
quantum field theory~\cite{Husain:2012im}, thermal effects in QFT~\cite{FS9506,Scardigli:2018jlm},
and for various deformations of the quantization rules~\cite{Jizba:2009qf}.
%

\section{Tests of GUP: Deformed Quantum Mechanics}
\vspace{2mm}

In the last decade a lively debate emerged
on the \emph{measurable\/} features predicted by the various kinds of Generalized Uncertainty
Principles (GUPs).
The discussion focused on the experimental predictions about the \emph{size\/} of these modifications, and among the better elaborated experimental proposals are those, for example, of the groups of Brukner, and Marin~\cite{bruk,bawaj}.
\par
We can roughly divide in two groups the Studies on the dimensionless deforming parameter $\beta$ of GUP .

In the first group, authors as Kempf, Mann, Brau, Vagenas, Nozari, etc. (Refs.\cite{KMM,brau,vagenas,Nozari}), translated the GUP into a deformed commutator and developed a deformed quantum mechanics. The deformed commutator 
(\ref{gupcomm}) is in general rewritten as
\be
\left[\hat{X},\hat{P}\right] = i\,\hbar\left(1 + \beta\, \frac{\hat{P}^2}{m_p^2 c^2}\right)
\ ,
\label{[1]}
\ee
where the fundamental variables $\hat{X}, \hat{P}$ are thought to be high energy operators valid, in particular, at or near the Planck scale. They have non linear representations, $\hat{X}=X(\hat{x})$, $\hat{P}=P(\hat{p})$ in terms of the operators $\hat{x}$, $\hat{p}$ which are the usual position and momentum operators at low energy, obeying the standard Heisenberg commutator 
$[\hat{x}, \hat{p}]=i\hbar$.

This approach is used in general to extract bounds on $\beta$ of non-gravitational origin by computing with the new $\hat{X}, \hat{P}$ well know physical phenomena, and comparing the results with experimental data. Explicit calculations rely on the specific transformation $\hat{X}=X(\hat{x})$, $\hat{P}=P(\hat{p})$, which is in general non linear, and also by definition non canonical, since $[\hat{x}, \hat{p}] \neq [\hat{X}, \hat{P}]$, namely the correspondent Poisson brackets are not preserved.

Authors compute corrections to quantum mechanical predictions, such as
energy shifts in the spectrum of the hydrogen atom, or to the Lamb shift,
the Landau levels, Scanning Tunneling Microscope, charmonium levels, etc.
As we shall see, the bounds on $\beta$ so obtained 
are quite stringent, but a debated drawback of
this approach is a possible dependence of the expected shifts on
the specific representation of the variables $X$ and $P$ in the
fundamental commutator~(\ref{[1]}).

\subsection{\textbf{Non gravitational bounds on $\beta$}}
\vspace{2mm}

The following are the the most commonly discussed (atomic) phenomena, where, by applying the above discussed Deformed Quantum Mechanics, non gravitational bounds on $\beta$ can be obtained.\\
\\
\noi $\bullet$ The Landau levels are recomputed with a GUP-corrected hamiltonian of the form
\be
H = \frac{1}{2m}(\vec{p_0} - e\vec{A})^2 + \frac{\beta_0}{m}(\vec{p_0} - e\vec{A})^4 = 
H_0 + 4\beta_0 m H_0^2
\ee
where $\beta_0=\beta/(m_p c)^2$, and for an electron in a magnetic field of $10 T$, $\omega_c\simeq 10^3$ GHz, 
authors in Refs.\cite{vagenas} (see also references therein) found a relative shift of the first level as
\be
\frac{\De E_{1(GUP)}}{E_1^{(0)}} \simeq 2.30 \times 10^{-54} \beta \,.
\ee
Since we have an accuracy of one part in $10^4$ in direct
measurements of Landau levels using a scanning tunnel
microscope (STM),
the upper bound on $\beta$ becomes
\be
10^{-54} \beta < 10^{-4} \Rightarrow \beta < 10^{50}\,.
\ee
\\
\noi$\bullet$ For the Lamb shift, again in Refs.\cite{vagenas}, authors get the a relative correction of
\be
\frac{\De E_{0(GUP)}}{\De E_0} \simeq 0.47 \times 10^{-48} \beta
\ee  
and since the current accuracy in precision measurements of Lamb shift is of about 1 part in 
$10^{12}$, this sets an upper bound of 
\be
\beta < 10^{36}.
\ee
\\
\noi$\bullet$ After having computed similar GUP-corrections for the harmonic oscillator, it is then possible to use them for systems that can be well modeled by the harmonic oscillator itself, such as heavy mesons like charmonium. In so doing the correction due to GUP can be calculated at the first order in $\beta$, and in Refs.\cite{vagenas} authors found for the relative shift in energy
\be
\frac{\De E_0^{(2)}}{E_0^{(0)}} \simeq 2.7 \times 10^{-39} \beta
\ee
The current accuracy in precision measurements for the case of $J/\psi$ is at the level of 
$10^{-5}$. Therefore they find an upper bound on $\beta$ as 
\be
\beta < 10^{34}.
\ee
\\
\noi$\bullet$ We then have bounds obtained without relying on a representation
of the fundamental commutator, but rather based directly on the deformed uncertainty
relations.
The first is coming from the lack of observed deviations from the standard model theory
at the electroweak scale (100~GeV).
In fact, Eq.~(\ref{gup}) can be promptly cast in the form
\be
\Delta x\, \Delta p
\geq
\frac{\hbar}{2}\left[ 1 + \beta \left(\frac{E_{\rm e w}}{E_p} \right)^2\right]
\ .
\ee
The absence of deviations from the standard Heisenberg principle at the electroweak
scale then implies $\beta(E_{\rm e w}/E_p)^2~\ll~1$, which means $\beta \ll 10^{34}$.\\
\\
\noi $\bullet$ The study of Bawaj et al.~\cite{bawaj} opened the way of probing deformed commutators with macroscopic harmonic oscillators. The authors have checked with different devices the behavior of state-of-the-art micro and nano mechanical oscillators. 

Supposing $m$ and $\omega_0$ are the mass and the frequency of the harmonic oscillator, then 
the classical standard Hamiltonian is $H=p^2/2m + m\omega_0^2 x^2/2 = \hbar \omega_0(Q^2 + P^2)/2$, where 
 $x=\sqrt{\hbar/m\omega_0}Q$, 
$p=\sqrt{\hbar m\omega_0}P$, with $Q$ and $P$ dimensionless coordinates. The operators $\hat{x}$ and $\hat{p}$ obey Eq.(\ref{gupcomm}), and therefore 
$[\hat{Q},\hat{P}]=i(1+\beta_0 \hat{P}^2)$ where here $\beta_0=\beta(\hbar m \omega_0/m_p^2 c^2)$. We now introduce the transformation $Q=Q$, $P=P(\tilde{P})$, where $\tilde{P}$ is the low energy momentum, such that $[Q,\tilde{P}]=i$. Therefore at the first order in $\beta_0$ we have 
$P=\tilde{P}(1+\beta_0\tilde{P}^2/3)$ and  
the GUP-deformed Hamiltonian for the harmonic oscillator reads
\be
H = \frac{\hbar \omega_0}{2}(Q^2 + \tilde{P}^2) + \frac{\hbar \omega_0}{3} \beta_0 \tilde{P}^4 \,.
\label{hambav}
\ee
The solution to the first order in $\beta_0$ of the equations of motion for such oscillator reads 
\be
Q=Q_0\left[\sin(\tilde{\omega}t) + \frac{\beta_0}{8}Q_0^2\sin(3\tilde{\omega}t) \right]
\ee
where $Q_0$ is the oscillation amplitude for $Q$ and
\be
\tilde{\omega} = (1+\frac{\beta_0}{2}Q_0^2)\omega_0 \,.
\ee 
Clearly, it presents two striking new effects with respect to the standard harmonic oscillator: a dependence of the oscillation frequency on the amplitude, and the appearance of a third harmonic.

Authors have examined three kinds of
oscillators, with masses of respectively $10^{-4}$, $10^{-7}$ and $10^{-11}$ kg. 
Results are summarized in Fig.1 (taken from Ref.\cite{bawaj}), where they are also compared with other bounds on $\beta$ listed above, or discussed in the next section.
\begin{figure}[t]
 \centering
 \includegraphics[width=15cm]{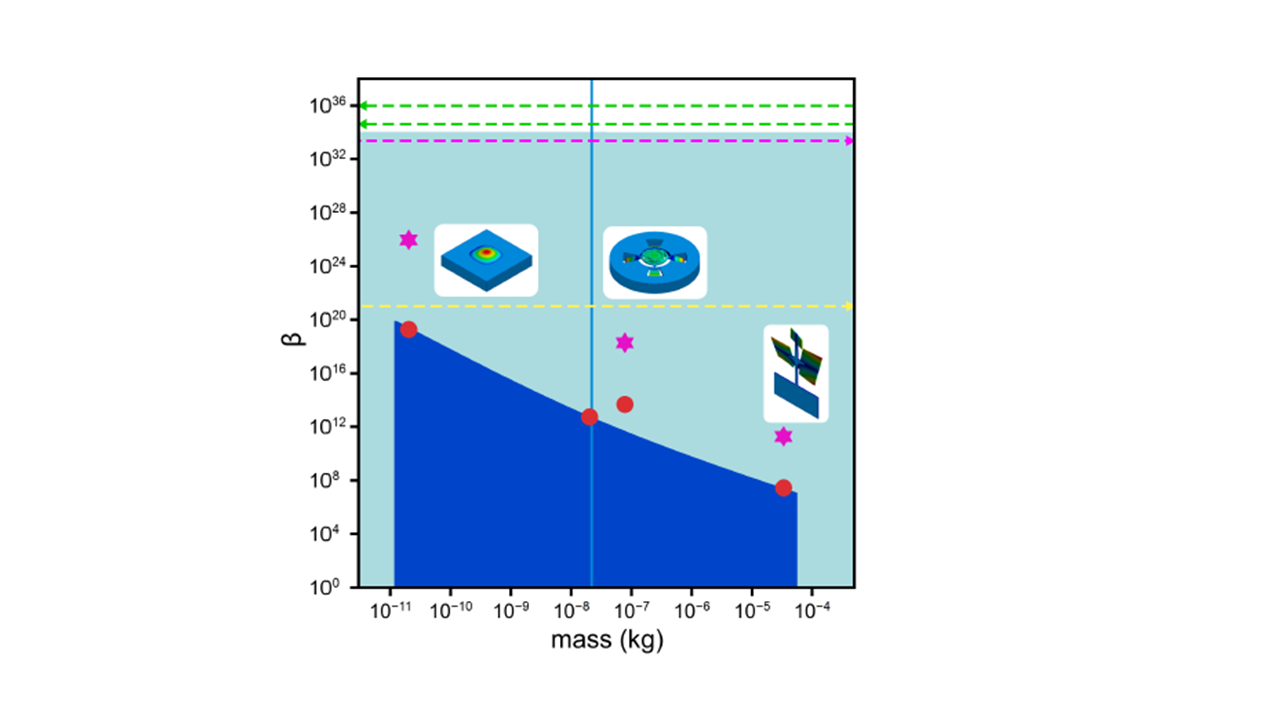}
 \renewcommand{\baselinestretch}{1.0}
 \caption{See comment in the text.}
 \label{figura}
 \end{figure}
The parameter $\beta$
quantifies the deformation to the standard commutator between position
and momentum. Its upper limits are reported on the Y-axis as a function of the mass involved in the measurement. 
Red dots come from the dependence of the oscillation frequency from its amplitude. They provide the best estimates of $\beta$ ranging from $10^8$ to $10^{19}$, with an average around $10^{12}$.
Magenta stars come from the third harmonic distortion, not present in the standard harmonic oscillator. Light blue shows the area below the electroweak
scale, dark blue the area that remains unexplored. Dashed
lines reports some previously estimated upper limits, obtained in
mass ranges outside this graph (as indicated by the arrows). Green:
from high resolution spectroscopy on the hydrogen atom, considering
the ground state Lamb shift (upper line) \cite{vagenas} and the 1S-2S level
difference (lower line) \cite{tcachuk}; on this lower green line there are also the bounds coming 
from the charmonium levels. Magenta: from the AURIGA detector
\cite{cerd}. Yellow: from the lack of violation of the equivalence principle
\cite{ghosh}. The vertical line corresponds to the Planck mass (22 $\mu$g).\\
\\
\noi $\bullet$ The very recent work of P.Bushev et al.~\cite{bushev} is a further brilliant step in the direction of the evaluation of $\beta$ with macroscopic harmonic oscillators. Actually, the sapphire mechanical resonator employed by Bushev is really "big", when compared to the standard devices used in in this field, in fact its mass amount to $0.3$ kg. The Hamiltonian governing the system is essentially again given by Eq.(\ref{hambav}), 
\be
H = \frac{\hat{p}^2}{2m} + \frac{1}{2}m\Omega_0^2 \hat{x}^2 + \frac{\beta \hat{p}^4}{3m (m_p c)^2}
\,, 
\ee
namely a deformation in $\beta\hat{p}^4$ of the usual harmonic oscillator Hamiltonian. Experiments with the sapphire resonator have established an upper limit on $\beta$ not exceeding 
$5.2 \times 10^6$, which is from 2 to 6 order of magnitude better than the one previously detected by Bawaj et al.. Moreover, Bushev's group is preparing a further experiment with a state of the art quartz BAW resonator, and they claim to have reasonable estimates that this would yield 
an even more stringent limit of $\beta < 4\times 10^4$.\\
\\
\noi $\bullet$ In Ref.~\cite{villa} authors deal with the broadening times of large molecular wave-packets (such as $C_{60}$, $C_{176}$ and large organic molecules) and the profound effect that a modified fundamental commutator should have on the dispersion rate of free wave-packets. Authors  propose experiments where the minimal length effect can be tested by measuring the timescale in which the wave-packet (associated with a particle or a system of particles) doubles its initial width. They start from a fundamental commutator containing also a linear term in $\hat{p}$ (see~\cite{ADV})
\be
[\hat{x},\hat{p}]=i\hbar(1-2\alpha_0 \hat{p} + 4\alpha_0^2 \hat{p}^2)
\ee
where $\alpha_0=\alpha/(m_p c)$. On comparing this with Eq.(\ref{gupcomm}), we see that 
$\beta \sim \alpha^2$.
The difference between the "doubling time" predicted by the HUP and the doubling time predicted by the GUP can be computed for a generic $\alpha$ and compared with experimental data. It turns out that $t_{double}$ is proportional to the mass $m$ of the particle considered, and that is why it is convenient to use large molecules. According to the authors, the use of organic macro molecules as the TPPF152 (consisting of 430 atoms) and of up to date atomic clocks (with an assumed precision of $10^{-15}$ s) would allow, in the proposed experiments, to probe values of $\alpha \sim 10^6$. Therefore this would give an upper bound for $\beta<10^{12}$.

\section{Tests of GUP: Deformed Classical Mechanics; gravitational bounds on $\beta$}
\vspace{2mm}

\par
In a second group of works 
(see, e.g., Chang~\cite{LNChang}
), classical Newtonian mechanics
is deformed by modifying the standard Poisson brackets in a way that resembles
the quantum commutator,
\be
\left[\hat{x},\hat{p}\right]
=
i\,\hbar\left(1 + \beta_0\, \hat{p}^2\right)
\quad
\Rightarrow
\quad
\{X,P\} = \left(1 + \beta_0\, P^2\right)
\ ,
\ee
where $\beta_0=\beta/(m_p^2 c^2)$.
In particular, in Ref.~\cite{LNChang} authors compute the precession of the
perihelion of Mercury directly from this GUP-deformed Newtonian mechanics,
and interpret it as an extra contribution to the well known precession of
$43"/$century due to General Relativity (GR).
They then compare this global result with the observational data,
and the very accurate agreement between the GR prediction and observations
leave them not much room for possible extra contributions to the precession.
In fact, they obtain the tremendously small bound $\beta < 10^{-66}$.
A problem with this approach is that this GUP-deformed Newtonian mechanics
is simply superposed linearly to the usual GR theory.
One may argue that a modification of GR at order $\beta$ should likewise
be considered, but this is however omitted in Ref.~\cite{LNChang}.
In other words, it is not clear why the two structures, GR and GUP-modified Newtonian
mechanics, should coexist independently, and why the two different precession errors
add into a final single precession angle.
Most important, as a matter of fact, in the limit $\beta \to 0$, Ref.~\cite{LNChang}
recovers {\em only} the Newtonian mechanics but not GR, and GR corrections must
be added as an extra structure.
Clearly, the physical relevance of this approach and the bound that follows
for $\beta$, remain therefore quite questionable.
\par
Finally, there are works on the evaluation of $\beta$ where authors (see for example 
Ghosh~\cite{ghosh}), in a similar but slightly different way, use a covariant formalism (first defined in Minkowski space, with the metric $\eta_{\mu\nu}={\rm diag}(1,-1,-1,-1)$, and then easily generalized to curved space-times via the standard procedure $\eta_{\mu\nu} \to g_{\mu\nu}$),
to introduce a deformation of classical Poisson brackets, although posited in covariant form.
From deformed covariant Poisson brackets, they obtain a $\beta$-deformed geodesic equation, which leads to a violation of the Equivalence Principle.
They do not deform the field equations or the metric.
It is however possible to show that this violation of the Equivalence Principle is
completely due to the postulate of deformed Poisson brackets, and has nothing to do with the general 
covariance of formalism, or with a deformation of the GR field equations or solutions. 
In other words, already a simple deformation of classical Newtonian
(i.e.~non covariant) Poisson brackets implies automatically a violation of the equivalence principle.
\par
In fact, suppose to deform 
Poisson brackets in the same fashion as the quantum commutators derived from the GUP 
(see Refs.~\cite{ghosh,LNChang}).
Then, considering just a one dimensional system to keep the calculation
simple, we can write the Poisson brackets of a pair of canonical variables
as
\be
\{q,p\} = 1 + \beta_0 \,p^2 \ ,
\quad \quad \quad
\{q,q\}=\{p,p\}=0
\ .
\label{mP}
\ee
Then for any regular function $H(q,p)$,
the following equations hold
\be
\{q,H\} = (1 + \beta_0 \,p^2)\frac{\partial H}{\partial p}
\ ,\quad\quad
\{p,H\} = -(1 + \beta_0 \,p^2)\frac{\partial H}{\partial q}
\ .
\ee
A point-like particle of mass $m$ moving in a Newtonian potential is described by
the Hamiltonian (assuming $M \gg m$)
\be
H = \frac{p^2}{2\,m} - \frac{G\,M\,m}{q}
\ ,
\ee
and evolves according to the equations of motion
\be
\dot{q}
=
\{q,H\}
=(1 + \beta_0\, p^2)\frac{p}{m}\, ,
\quad\quad\quad
\label{eqm}
\\
\dot{p}
=
\{p,H\}
=
-(1 + \beta_0\, p^2)\frac{G\,M\,m}{q^2}
\ .
\ee
To first order in $\beta$, we have therefore the equation of motion
\be
\ddot{q}
\simeq
- (1 + 4\,\beta_0\,(m\,\dot{q})^2)\,\frac{G\,M}{q^2}
\ .
\ee
Clearly, the trajectory of a test particle of mass $m$ will depend on $m$,
which signals a violation of the Equivalence Principle (see Refs.~\cite{SC,AFA}).
This violation has nothing to do with covariant formalism, GR, or the geodesic equation, but strictly follows, even in simple Newtonian mechanics, from the modified Poisson brackets~(\ref{mP}).

However, the Ghosh formalism remains covariant when $\beta \to 0$
and reproduces standard GR results in the limit $\beta \to 0$.
In conclusion, the bound on $\beta$ predicted by Ghosh (see Ref.~\cite{ghosh}) from a possible violation of the universality of free fall, namely, from a violation of the (weak) equivalence principle, is $\beta < 10^{21}$.

The role of the equivalence principle in the GUP context has been discussed also in
Ref.~\cite{tkachuk2012}. 
There, Tkachuk (2012) recovers Equivalence Principle also for deformed classical
mechanics (i.e. deformed Poisson brackets) by considering composite bodies, and
postulating that kinetic energy has the additive property, that is, it does not depend
on the composition of a body but only on its total mass. The price to pay is to introduce a deformation parameter $\beta_{0i}$ which depends on the mass $m_i$ of the i-th (elementary) particle composing the macroscopic body. In other words, each (elementary) particle composing the body has its own deformation parameter $\beta_{0i}$. Then a new universal constant is introduced, 
$\gamma^2 =\beta_{0i} m_i^2$, which allows to recover the validity of the equivalence principle for deformed classical mechanics (by the way, to the first order in $\beta$). The construction is without doubt ingenious, however has the drawback that each specific particle would have its own specific minimal length, $\hbar \sqrt{\beta_{0i}} = \ell_i$. The effective minimal length that can be probed by a proton should be different (specifically, smaller) from that of an electron. This feature clearly seems to be at odd with the universality of gravitation, and with the fact that  Planck length can be computed in a way completely independent from the particle considered 
(see e.g.~\cite{FS9506}).     
%
%
\subsection{\textbf{Gravitational bounds on $\beta$ preserving the equivalence principle}}
\vspace{2mm}

By exploiting the (first) gravitational wave event ever observed, GW150914, authors of 
Ref.~\cite{Feng} have been able to set bounds on $\beta$ of purely gravitational origin. They start from a representation of the deformed commutator $[X_i,P_j]=i\hbar\delta_{ij}(1+\beta_0P^2)$
which reads
\be
X_i=x_{0i} \, , \quad \quad \quad P_i=p_{0i}(1+\beta_0p^2)
\ee
where $x_{0i}$, $p_{0i}$ are the low energy position and momentum operators, satisfying the canonical commutation relations $[x_{0i}, p_{0j} ] = i\hbar \delta_{ij}$, 
and $p^2 = \eta_{ij}p^{0i}p^{0j}$. From here authors deduce a GUP-modified dispersion relation (to the first order in $\beta$) 
\be
E^2=m^2c^4 + p^2c^2(1-2\beta_0p^2)\,.
\ee
Therefore the group speed for massless gravitons will be
\be
v_{massless} = \frac{\partial E}{\partial p} \simeq c(1 -3\beta_0 p^2)\,.
\label{speed}
\ee
Then, assuming for massless gravitons $E_g \simeq p_g c$, they can write 
$v_{massless} \simeq c(1 -3\beta_0 E_g^2/c^2)$ where $E_g$ is the energy of gravitons. Finally, the difference between the modified speed of gravitons and the speed of light is
\be
\Delta v = c - v_{massless} = 3 \beta \frac{E_g^2}{m_p^2c^3}\,.
\ee
From the data of GW150914 we get an upper bound
\be
3 \beta \frac{E_g^2}{m_p^2c^3} < 5.6 \times 10^{-12} \quad m/s \,.
\ee
Since the signal of the gravitational wave event GW150914 was peaked at 150Hz, 
here they can write $E_g \simeq 6.024 \times 10^{-13}$eV.
Therefore, the upper bound on the GUP parameter $\beta$ turns out to be
\be
\beta < 2.3 \times 10^{60} \,,
\ee
which is more stringent than the gravitational bounds derived, for example, in Ref.\cite{SC}.
We emphasize that in this procedure the authors, because of GUP, have modified the usual dispersion relation. Namely the propagation speed of gravitons is not the same for all the frequencies/energies, but depends on the frequencies/energies of the gravitons considered, according to 
Eq.(\ref{speed}). Note however that this procedure does not violate the Equivalence Principle (EP). In fact, we have a violation of the EP when the equation of motion contains deforming terms dependent on $\beta$ and/or on the mass $m$ of the test body, and therefore such equation does not coincide anymore with the geodesic equation. (The geodesic equation is given and fixed, once the metric is given.) In cases in which the trajectory of a test body depends on the mass of the test body, then we have a violation of EP. 
But of course, quite obviously, a generic trajectory/geodesic should depend on the (initial) velocity of the moving body. For example, the trajectory of a planet around the Sun ($V_{planet}\ll c$) is not the trajectory of a photon in the gravitational field of the Sun ($V_{photon}=c$). Nevertheless both trajectories are geodesics of the metric surrounding the Sun. Hence, a dispersion relation requiring different speeds for different frequencies, will imply that red light follows a path different from blue light, but however both paths will be geodesics of the metric (not trajectories different from geodesics). Therefore the EP is not violated.\\
\\
A further example of bounds on $\beta$ of gravitational origin, but not depending on a violation of the EP, is given in Ref.\cite{SC}. There, authors consider a deformation of the Hawking temperature of a Schwarzschild black hole when computed through the GUP. The GUP deformed Hawking temperature turns out to be (at first order in $\beta$)
\be
T(\beta) = \frac{\hbar c^3}{8\pi G k_B M}\left(1 + \frac{\beta m_p^2}{4\pi^2 M^2} + \dots \right)\,.
\label{tbeta}
\ee       
Such deformed temperature can be mimicked through the standard Hawking temperature of a deformed Schwarzschild black hole metric. In fact, authors show that a deformed spherically symmetric metric like
\be
ds^2 = F(r)dt^2 - F(r)^{-1}dr^2 - r^2 d\Omega^2
\ee
with
\be
F(r) = 1 - \frac{2GM}{r c^2} + \varepsilon \frac{G^2 M^2}{r^2 c^4}
\label{dm}
\ee
has an horizon radius $r_H = (GM/c^2)(1+\sqrt{1-\varepsilon})$, and the correspondent standard Hawking temperature is 
\be
T(\varepsilon) = \frac{\hbar c }{4\pi k_B}F'(r_H) = 
\frac{\hbar c^3}{2\pi G k_B M}\frac{\sqrt{1-\varepsilon}}{(1+\sqrt{1-\varepsilon})^2}\,.
\ee
Imposing $T(\beta) = T(\varepsilon)$, they get a relation between $\beta$ and $\varepsilon$ that at the lowest order reads
\be
\beta = - \frac{\pi^2 M^2 }{4 m_p^2} \varepsilon^2 \,.
\label{be}
\ee
Once we have a deformed metric, we can use very precise astronomical measures to set bounds on the magnitude of the deformation $\varepsilon$, and then transfer these bounds on $\beta$, through Eq.(\ref{be}).

Incidentally, we note that the negative value of the $\beta$ parameter predicted by Eq.(\ref{be}), although it may appear surprising, has been already encountered in different scenarios: for an uncertainty relation  formulated on a (crystal) lattice (see \cite{Jizba:2009qf}); in the Magueijo and Smolin formulation of Doubly Special Relativity \cite{MS}, where the fundamental commutator 
$[X,P]=i\hbar (1-E/E_p)$ exhibits a $\hbar$ that can be interpreted as depending on the energy, $\hbar(E) \to 0$ when $E \to E_{Planck}$; in the analysis of the Chandrasekhar limit \cite{Ong}, where Y.C.Ong shows that $\beta$ should be negative in order to have a match between GUP-predicted white dwarfs masses and the astrophysical measures; in the comparison between the modification of Hawking radiation predicted by GUP and that predicted by corpuscular models of Gravity~\cite{GPLPetruzz, Dvali, Cas}. All these hints point towards a negative $\beta$ which would implies, immediately, 
$[\hat{X},\hat{P}]=i\hbar (1 + \beta \hat{P}^2/m_p^2c^2) \to 0$ for $P \to P_{Planck}$. Namely, a negative $\beta$ implies the existence of a sharp, \textit{classical world} at the Planck Scale, where $[\hat{X},\hat{P}]=0$. This would realize the vision formulated by G.'t Hooft in his deterministic interpretation of Quantum Mechanics, where the world is a gigantic, deterministic cellular automaton which works \textit{CLASSICALLY} at the Planck scale \cite{tHooft}. \\  
\\
To constrain $\varepsilon$, the best data come from the measures of the precession of Mercury. Using the deformed metric (\ref{dm}), we compute that the total precession after a single lap is
\be
\Delta \phi = 6\pi \frac{GM}{L}\left(1 - \frac{\varepsilon}{6}\right)
\ee 
to the first order in $GM/L$ (where $L=(1-e^2)a$ is the \textit{semilatus rectum} of 
Mercuty orbit). Note that for $\varepsilon \to 0$ we recover the usual, well known, General Relativity prediction. The data provided by the Messenger spacecraft \cite{laskar}, which orbited Mercury for two years in 2011-2013, give us a bound on $\varepsilon$ as
\be
|\varepsilon| < 1.6 \times 10^{-4}
\ee  
which implies the following bound on $\beta$
\be
|\beta| < 2 \times 10^{69}\,.
\ee 
Although much worse than the $\beta < 10^{21}$ obtained by Ghosh, this bound has been obtained with a procedure that fully preserves the equivalence principle. 
%
%
\section{Theoretical calculation of $\beta$}
\vspace{2mm}

By following a path similar to the one just presented in the previous section (formulae 32-36), it possible to compute an \textit{exact} value of $\beta$ (see Ref.~\cite{SLV}). In fact, suppose to consider a generic deformation of the spherically symmetric metric given above, as 
\be
F(r) = 1 - \frac{2GM}{r c^2} + \varepsilon \phi(r)\,.
\label{defF}
\ee  
Then the deformed Schwarzschild radius (to the first order in $\varepsilon$) reads 
\be
r_H \ = \ a - \frac{\varepsilon\, a\, \phi(a)}{1 \ + \ \varepsilon\,[\phi(a) \ + \ a\,\phi'(a)]}\,,
\label{solr}
\ee
where $a = 2 G M/c^2$, and the corresponding deformed standard Hawking temperature is 
\be
T =  \frac{\hbar c }{4\pi k_B}F'(r_H) = \frac{\hbar c^3}{8\pi k_B GM} \left\{1 + \varepsilon \left[2\phi(a) + a\phi'(a)\right] 
+ \varepsilon^2 \phi(a) \left[\phi(a) - 2a\phi'(a) - a^2 \phi''(a) \right] + \dots \right\}.
\label{teps}
\ee
Now, supposing to know the exact form of the correction term $\varepsilon \phi(r) \equiv \epsilon(r)$ in formula (\ref{defF}), then we can compare the two expansions (\ref{teps}) and (\ref{tbeta}), and from the two respective first order terms, we can extract the exact value of $\beta$ as
\be
\beta \ = \  \frac{4\pi^2 M^2}{m_p^2}\, \left[2\epsilon(a) + a\epsilon'(a)\right]~.
\label{beta1}
\ee

This is what actually happens, when we consider the leading quantum corrections to the Newtonian potential computed, in an early attempt, by Duff~\cite{Duff}, and more recently by 
Donoghue~\cite{JD}.

Donoghue reformulated General Relativity as an effective field theory. At ordinary energies gravity is a well behaved QFT. The quantum corrections at low energy, and dominant
effects at large distances can be isolated, and are shown to be due to the
propagation of massless particles (gravitons). The effective quantum-corrected gravitational potential generated by an heavy mass $M$, interacting with another heavy body $m$ (both bodies close to rest), then reads
\be
V(r)=-\frac{GM}{r}\left(1 + \frac{3GM}{r c^2}(1+\frac{m}{M}) +  \frac{41}{10\pi}\frac{\ell_p^2}{r^2}\right) .
\label{DP}
\ee
The first correction term does not contain any power of $\hbar$, so it is a classical effect, due to the non-linear nature of General Relativity. However, the second correction term, i.e., the last term of  (\ref{DP}), is a true quantum effect, linear in $\hbar$.

Now, it is well know that for any metric of the form 
$ds^2 = F(r)dt^2 - F(r)^{-1}dr^2 - C(r)d\Omega^2$ an effective Newtonian potential \footnote{The effective Newtonian potential is produced by such metric on a point particle which moves slowly, in a stationary and weak gravitational field, i.e., quasi-Minkowskian far from the source, $r \to \infty$.} (i.e. the Newtonian limit) can be defined as
\be
V(r) \ \simeq \ \frac{c^2}{2} \ (F(r)-1)
\ee
Conversely, a metric mimicking a given Newtonian potential $V(r)$ will be
\be
F(r) \ \simeq \ 1 \ + \ 2 \ \frac{V(r)}{c^2}~.
\ee
Therefore, the metric mimicking the quantum-corrected Newtonian potential (\ref{DP}) will be
\be
F(r) \ \simeq \ 1 + 2 \frac{V(r)}{c^2} \ =  
1 - \frac{2GM}{r c^2} - \frac{6\, G^2 M^2}{r^2 c^4}\left(1+\frac{m}{M}\right) -
\frac{41}{5\pi}\frac{G^3 M^3}{r^3 c^6}\left(\frac{\ell_p\, c^2}{GM}\right)^{2}.\nonumber
\ee
Identifying now
\be
\epsilon(r) = - \frac{6\, G^2 M^2}{r^2 c^4}\left(1+\frac{m}{M}\right) -
\frac{41}{5\pi}\frac{G^3 M^3}{r^3 c^6}\left(\frac{\ell_p\, c^2}{GM}\right)^{2}\,,
\ee
then, through Eq.(\ref{beta1}), we finally arrive at
\be
\beta = \frac{4\pi^2 M^2}{m_p^2}\,\frac{41}{40 \pi}\left(\frac{\ell_p c^2}{GM}\right)^2
= \frac{82\pi}{5}~,
\ee
where we used $\ell_p c^2/GM = 2m_p/M$.
Therefore, $\beta$ results to be roughly of order $1$, as expected also from other completely different approaches (among the many, see for example \cite{VenezGrossMende}, \cite{MaxAcc}).
%
%
\section{Summary and conclusions}
\vspace{2mm}
The main results described in this paper are about the bounds and the value of the deformation parameter $\beta$ of the GUP, and they can be summarized as follows.\\
\\
\noi $\bullet$ The best bound on $\beta$ of non-gravitational origin  seems to be in the range $\beta < 10^{8} \div 10^{12}$ (Bawaj, Marin 2015), or probably even $\beta < 5\times 10^6$ 
(Bushev 2019)\\
\\
\noi $\bullet$ The best bound on $\beta$ of gravitational origin (if we allow for a violation of the  Equivalence Principle)  seems to be $\beta < 10^{21}$ (Ghosh 2014).\\
\\
\noi $\bullet$ The best bound on $\beta$ of gravitational origin (preserving however the Equivalence Principle) seems to be $\beta < 10^{60}$ (Feng 2017).\\
\\
\noi $\bullet$ Different theoretical frameworks, and an explicit calculation (Ref.\cite{SLV}), indicate a value of $\beta$ of order 1, in particular $\beta \simeq 82\pi/5$.\\
\\
So we can conclude that the gap between theoretical predictions and experimental bounds will still require a quite big leap in experimental techniques in order to probe a vast region for the parameter $\beta$. For sure, there is a lot of space for improvement!

\section*{References}

\end{document}